\documentclass[12pt]{article}
\usepackage{graphicx}
\usepackage{subfigure}
\newcommand{\be}{\begin{equation}}
\newcommand{\ee}{\end{equation}}

\newcommand{\beqa}{\begin{eqnarray}}
\newcommand{\eeqa}{\end{eqnarray}}

\newcommand{\eqref}[1]{(\ref{#1})}

\def\boxit#1{\vbox{\hrule\hbox{\vrule\kern8pt
\vbox{\hbox{\kern8pt}\hbox{\vbox{#1}}\hbox{\kern8pt}}
\kern8pt\vrule}\hrule}}
\def\mathboxit#1{\vbox{\hrule\hbox{\vrule\kern8pt\vbox{\kern8pt
\hbox{$\displaystyle #1$}\kern8pt}\kern8pt\vrule}\hrule}}

\def\IB{\relax\hbox{$\inbar\kern-.3em{\rm B}$}}
\def\IC{\relax\hbox{$\inbar\kern-.3em{\rm C}$}}
\def\ID{\relax\hbox{$\inbar\kern-.3em{\rm D}$}}
\def\IE{\relax\hbox{$\inbar\kern-.3em{\rm E}$}}
\def\IF{\relax\hbox{$\inbar\kern-.3em{\rm F}$}}
\def\IG{\relax\hbox{$\inbar\kern-.3em{\rm G}$}}
\def\IGa{\relax\hbox{${\rm I}\kern-.18em\Gamma$}}
\def\IH{\relax{\rm I\kern-.18em H}}
\def\IK{\relax{\rm I\kern-.18em K}}
\def\IL{\relax{\rm I\kern-.18em L}}
\def\IP{\relax{\rm I\kern-.18em P}}
\def\IR{\relax{\rm I\kern-.18em R}}
\def\IZ{\relax\ifmmode\mathchoice
{\hbox{\cmss Z\kern-.4em Z}}{\hbox{\cmss Z\kern-.4em Z}}
{\lower.9pt\hbox{\cmsss Z\kern-.4em Z}} {\lower1.2pt\hbox{\cmsss
Z\kern-.4em Z}}\else{\cmss Z\kern-.4em Z}\fi}

\def\II{\relax{\rm I\kern-.18em I}}

\begin{document}

\thispagestyle{empty}

\vspace*{2.5cm}

\begin{center}

{\large {\bf Three-dimensional Gonihedric Potts model}}

\vspace*{2.5cm}

{\bf P.Dimopoulos}\\

INFN--ROME2, \\

Dipartimento di Fisica, Universita di Roma 'Tor Vergata', \\

Via della Ricerca Scientifica, Roma, Italy\\

email:dimopoulos@roma2.infn.it

\vspace{1cm}

{\bf G.Koutsoumbas}\\

Physics Department, National Technical University, \\

Zografou Campus, 15780 Athens, Greece\\

email:kutsubas@central.ntua.gr

\vspace{1cm}

{\bf G.K.Savvidy}\\

National Research Center Demokritos,\\

Ag. Paraskevi, GR-15310 Athens, Greece\\

email:savvidy@inp.demokritos.gr

\end{center}

\vspace{60pt}

\centerline{{\bf Abstract}}

\vspace{12pt} \noindent We study, by the Mean Field and Monte
Carlo methods, a generalized q-state Potts gonihedric model. The
phase transition of the model becomes stronger with increasing
$q.$ The value $k_c(q),$ at which the phase transition becomes
second order, turns out to be an increasing function of $q.$

\newpage

\section{Introduction}

In recent articles
\cite{pav}--\cite{george} the authors considered a class of models
describing two-dimensional random surfaces embedded into an
Euclidean lattice $Z^d,$ where a closed surface is associated with
a collection of plaquettes. This class of models is defined by
their statistical weights which are proportional to the total
number of non-flat edges of the surface \cite{sav1}, that {\it to
the extrinsic curvature of the surface}.\\

Various models of random surfaces built out of plaquettes, but
with different geometrical weights have been considered earlier in
the literature \cite{weingarten} and they describe random surfaces
with the Boltzmann weights which are proportional to the total
number of plaquettes
\cite{weg}, that is {\it to the area of the surface}.\\

Statistical and critical properties of these diverse classes
of models: the one which describes random surfaces with area action
and the one with extrinsic curvature action,
have been a subject of intensive
studies in the recent years.  The main interest is to
learn more about their physical properties and to understand
to what extent the geometrical nature of these models
spell out their critical behavior.\\

Before defining the generalization of the Potts model, which we
shall build using the geometrical concept of extrinsic curvature,
let us first recall the definition of the gonihedric model of
random surfaces  with extrinsic curvature action \cite{sav1,sav}.
As we just mention above it corresponds to a statistical system
with weights proportional to the total number of non-flat edges
$n_2$ of the surface. The weights associated with
self-intersections are proportional to $k n_4$ where $n_4$ is the
number of edges with four intersecting plaquettes, and $k$ is the
self-intersection coupling constant \cite{sav1,sav}. The partition
function is a sum over two-dimensional surfaces of the type
described above, embedded in a three-dimensional lattice: \be
Z(\beta) = \sum_{\{surfaces~M\}} e^{-\beta~\epsilon(M)},
\label{partfan} \ee where $\epsilon(M)=n_2 + 4 k n_4$ is the
energy of the surface $M$. In three dimensions the equivalent spin
Hamiltonian is equal to \cite{sav} \be\label{gonihedricham}
H_{gonihedric}^{3d}=- 2k \sum_{\vec{r},\vec{\alpha}}
\sigma_{\vec{r}} \sigma_{\vec{r}+\vec{\alpha}} + \frac{k}{2}
\sum_{\vec{r},\vec{\alpha},\vec{\beta}} \sigma_{\vec{r}}
\sigma_{\vec{r}+\vec{\alpha} +\vec{\beta}} -  \frac{1-k}{2}
\sum_{\vec{r},\vec{\alpha},\vec{\beta}} \sigma_{\vec{r}}
\sigma_{\vec{r}+\vec{\alpha}}
\sigma_{\vec{r}+\vec{\alpha}+\vec{\beta}}
\sigma_{\vec{r}+\vec{\beta}}, \label{hamil} \ee and it is an
alternative  model to the $3D$ Ising system \cite{weg}
$$
H_{Ising}^{3d}= -  \sum_{\vec{r},\vec{\alpha}} \sigma_{\vec{r}}
\sigma_{\vec{r}+\vec{\alpha}}
$$
for which the energy is proportional to the area
$\epsilon(M)=n_1$, where $n_1$ is the total number of plaquettes.\\

The degeneracy of the vacuum state depends on
self-intersection coupling constant $k$ \cite{pav}. If $k \neq 0$,
the degeneracy of the vacuum state is equal to $3\cdot 2^N$ for
the lattice of size $N^3,$ while it equals $2^{3N}$ when $k=0$.
The last case is a sort of supersymmetric point in the space of
gonihedric Hamiltonians \cite{pav}
\be
H_{gonihedric}^{k=0}=-\frac{1}{2}\sum_{\vec{r},\vec{\alpha},\vec{\beta}}
\sigma_{\vec{r}} \sigma_{\vec{r}+\vec{\alpha}}
\sigma_{\vec{r}+\vec{\alpha}+\vec{\beta}}
\sigma_{\vec{r}+\vec{\beta}}. \label{k=0case}
\ee
To study statistical and scaling properties of the system one can
directly simulate surfaces by gluing together plaquettes with the
corresponding weight $exp(-\beta (n_2 + 4 k n_4))$ or (much
easier) to study an equivalent spin system (\ref{gonihedricham}).
The first Monte Carlo simulations
\cite{bathas,des,cappi} demonstrate
that the gonihedric system
with intersection coupling constant greater than $k_c \approx 0.5$
(including $k=1),$ undergoes a second order phase transition at
$\beta_{c} \approx 0.44$ and that the critical indices are
different from those of the 3D Ising model.
Thus they are in
different classes of universality. On the contrary, the system
shows a first order phase transition for $k < k_c,$
including the ``supersymmetric" point $k=0$.\\

In the present work we would like to study a model which has a
Hamiltonian similar to the one defined by the equation
(\ref{gonihedricham}), but when the spin variable $\sigma$ runs
over larger field, in particular it takes on $q$ different values,
namely $\sigma = 0, 1, 2, \dots, q-1$. The basic principle we will
follow in this generalization is to keep the same geometrical
structure of the model. \\

\section{The model}

\vspace{.5cm}

The Hamiltonian of the model that we will study reads:

\begin{eqnarray}
H_{Gonihedric Potts}^{3d}&=&- 2 k \sum_{x,\mu>0} (2
\delta_{\sigma_x,\sigma_{x+\mu}}-1) + \frac{k}{2}\sum_{x,\mu \ne
\nu, \mu>0} (2 \delta_{\sigma_x,\sigma_{x+\mu+\nu}}-1) \nonumber
\\ &-& \frac{1-k}{2}\sum_{x,\mu \ne \nu, \mu>0, \nu>0} (2
\delta_{\sigma_x,\sigma_{x+\mu}}\delta_{\sigma_{x+\nu},
\sigma_{x+\mu+\nu}}-1) \cdot \nonumber \\ && \hspace*{3.2cm} (2
\delta_{\sigma_x,\sigma_{x+\nu}}\delta_{\sigma_{x+\mu},
\sigma_{x+\mu+\nu}}-1) \cdot \nonumber \\ && \hspace*{3.2cm}(2
\delta_{\sigma_x,\sigma_{x+\mu+\nu}} \delta_{\sigma_{x+\nu},
\sigma_{x+\mu}}-1) \label{hamil1}
\end{eqnarray}
The spin variables live on the sites of the lattice and take on
the values $0, 1, 2, \dots, q-1$ for the q-state Potts model. For
$q=2$ the model reduces to the standard gonihedric model. In
particular, the first two terms correspond to nearest neighbour
and next to nearest neighbour interactions, the third term
reproduces the $q=2$ plaquette interaction. Furthermore the
definition of the plaquette interaction in Eq.(\ref{hamil1}) also
describes a plaquette for $q>2$ in the sense that the  state with
the minimum energy  is reached at whenever the four spins are
pairwise equal.

Thus the model (\ref{hamil1}) becomes a generalization of the $3D$
Gonihedric Model. The degeneracy of the vacuum state depends on
the self-intersection coupling constant $k$ \cite{pav}. If $k \neq
0$, the degeneracy of the vacuum state is equal to $3\cdot q^N$
for the lattice of size $N^3,$ while it equals $q^{3N}$ when
$k=0$. The last case is again a sort of supersymmetric point in
the space of gonihedric Hamiltonians.

In Figure 1 one can see all possible surface configurations
appearing on a given edge for $\sigma = 0,\pm 1$ (we have here q=3
and plaquettes are drawn normal to the plane of the figure). As
one can see we have to include two new configurations whose energy
depends on the position of the third state. The corresponding
energies are equal to $\varepsilon_1 = 4k,~\varepsilon_2 =
0,~\varepsilon_3 = 1, ~\varepsilon_4 = 0,~\varepsilon_5 =
1,~\varepsilon_6 = 2k+1$. This model can be considered also as a
generalization of the Potts model. Relative to the gonihedric
models described above this new model has bigger entropy for the
given surface configuration, therefore one can qualitatively
predict that it should belong to the same universality class, but
the critical temperature should be higher. Indeed, it appears that
it has very similar phase structure but the critical temperature
and the critical intersection coupling constant are slightly
larger compared to the original model.

\begin{figure}
\begin{center}
\includegraphics[scale=0.80]{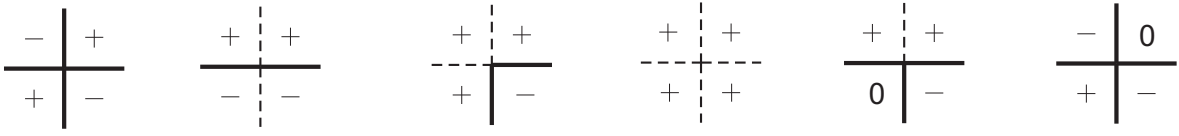}
\caption[fig0]{Different plaquette configurations on a given edge
for the case $\sigma = 0,\pm 1$ (here q=3). Plaquettes  and the
edges are drawn normal to the plane of the figure, the edge is at
the center of the intersections. The spin variables are on the
centers of the dual lattice. The corresponding energies are equal
to $\varepsilon_1 = 4k,~\varepsilon_2 = 0,~\varepsilon_3 = 1,
~\varepsilon_4 = 0,~\varepsilon_5 = 1,~\varepsilon_6 = 2k+1$ (from
left to right). Only non-flat edges contribute to the energy
functional $\varepsilon_3 = \varepsilon_5 =1$, the intersections
are suppressed by a factor proportional to intersection coupling
constant $k,$ $\varepsilon_1 = 4k,~\varepsilon_6 = 2k+1$.}
\label{fig0}
\end{center}
\end{figure}

\section{Mean field analysis}

It is useful to analyse our model using the mean field
approximation, before employing the rather expensive Monte Carlo
simulations. Since none of the $q$ states of the model is
preferable over the others, one may assume that the fractions
$x_n$ of the spins that take the value $n (n=0,\dots,q-1)$ may be
expressed in the form:
$$ x_0(s)=\frac{1+(q-1) s}{q},$$
\be
~~x_n(s)=\frac{1-s}{q}, n=1,\dots,q-1.
\ee
The order parameter $s$ will
be the one which minimizes the free energy. The value $s=0$
induces equal values to the $x_n(s)$ while for $s=1$ we get
$x_0(1)=1, x_n(1)=0,$ representing a long range order. With the
remark that for each lattice site we have 3 first neighbours, 6
second neighbours and 3 plaquettes, we may easily write down the
energy $E$ per site that enters in the mean field calculations:
\be E= (-2 k) \cdot 3 \cdot \sum_{n=0}^{q-1} x_n^2(s) +(\frac{k}{2})
\cdot 6 \cdot \sum_{n=0}^{q-1} x_n^2(s) +(-\frac{1-k}{2}) \cdot 3
\cdot \left( \frac{q \sum_{n=0}^{q-1} x_n^2(s) -1}{q-1}\right)^2
\label{en1} \ee \be =-6 k \sum_{n=0}^{q-1} x_n^2(s) + 3 k
\sum_{n=0}^{q-1} x_n^2(s) -\frac{3 (1-k)}{2} s^4 \label{en2} \ee The
sum representing the plaquette term has been written in the form
indicated in equation (\ref{en1}), so that a $s^4$ term is
reproduced. If  one used the form $\left(\sum_{n=0}^{q-1} x_n^2(s)
\right)^2$ instead, terms proportional to $s^2$ would also be
present, contaminating the nearest neighbour terms. The
calculations involves minimization of the free energy $$F \equiv
\beta E - S,$$ where the entropy $S$ is given by: $$S =
-\sum_{n=0}^{q-1} x_n(s) \ln (x_n(s)).$$

\begin{figure}
\begin{center}
\includegraphics[scale=0.40]{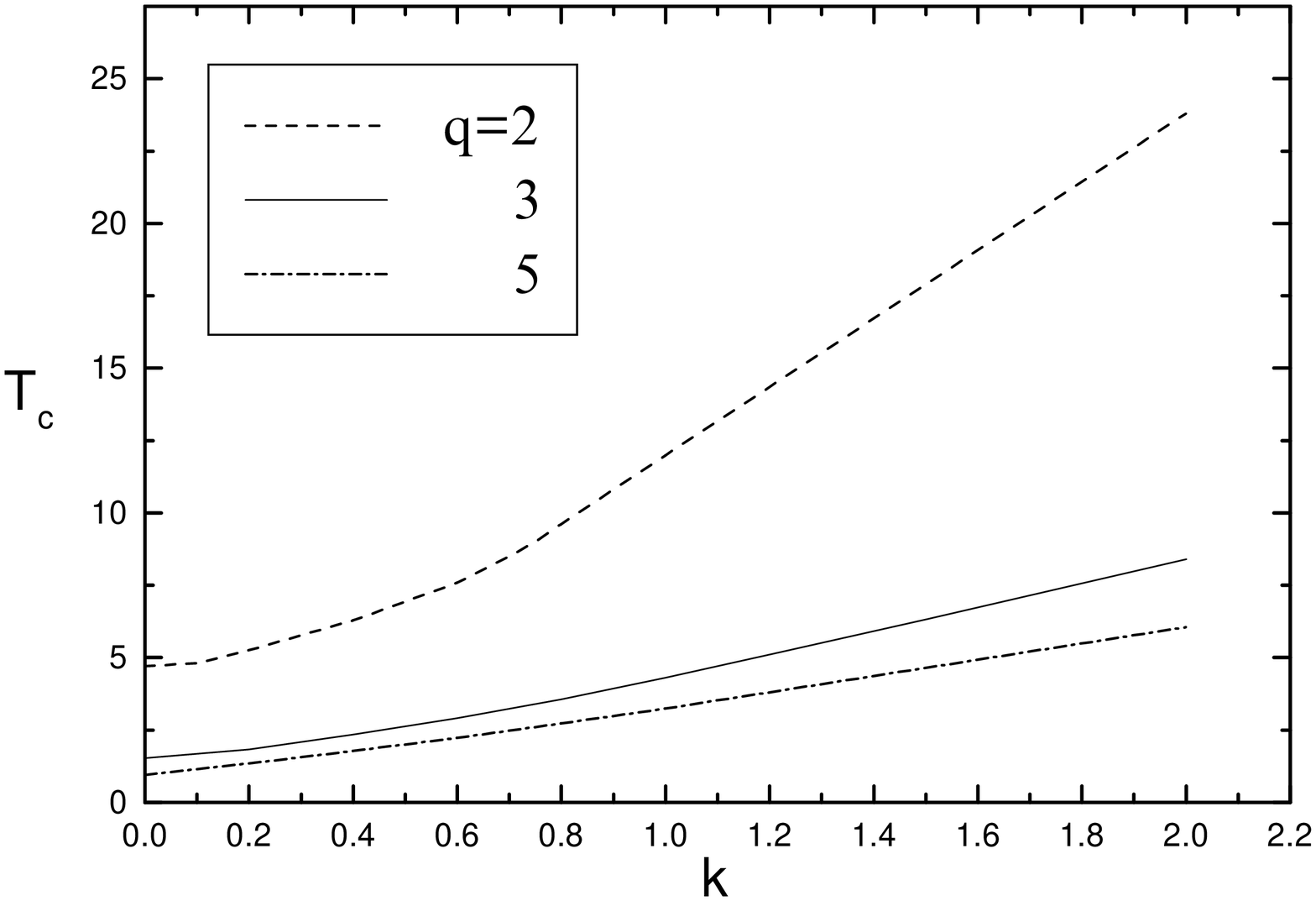}
\caption[tvk]{Mean Field result: Critical Temperature versus $k$ for various values
of $q.$} \label{fig1}
\end{center}
\end{figure}

In Fig. \ref{fig1} we show the mean field estimate of the critical
temperature as a function of $k$ for $q=2, 3, 5.$ We observe that
the critical temperature is highest (at the same value of $k$) for
$q=2.$ In general the critical temperature increases with $k.$


\begin{figure}
\begin{center}
\includegraphics[scale=0.40]{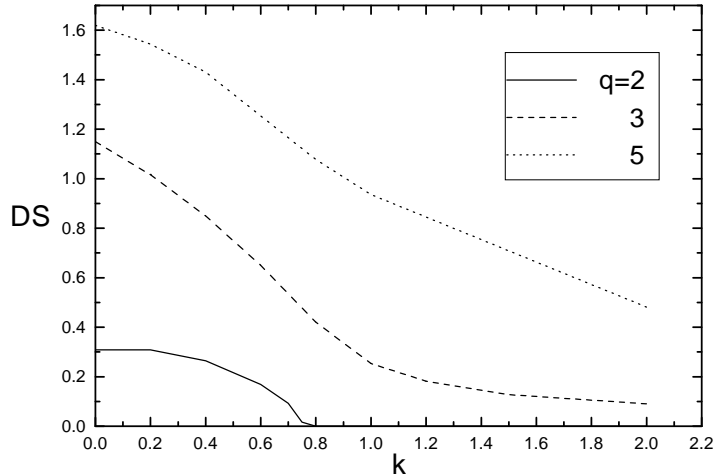}
\caption[svk]{Mean Field result: Latent Heat versus $k$ for various values of $q.$}
\label{fig2}
\end{center}
\end{figure}

A very interesting issue is the strength of the phase transition.
This is quantified by the latent heat, which is essentially the
entropy difference between the two phases accross the phase
transition. The mean field predictions are contained in
Fig.\ref{fig2}. The striking feature is that for $q=2$ the latent
heat starts from a (rather small) value and dives to zero as $k$
becomes bigger than about 0.8. Thus the prediction is that a first
order transition becomes second order at some value of $k.$ This
qualitatively agrees with known results about the standard
gonihedric model. We note that this change takes actually place at
$k \simeq 0.5,$ so the agreement is only approximate. This
characteristic is not preserved for $q>2.$ It is evident from the
figure that the latent heats for $q=3$ and $q=5$ are bigger than
the ones for $q=2.$ The most important fact, however, is that the
latent heats never become zero; instead they decrease monotonously
with $k$ and eventually, for large enough $k$ there will be a weak
first order transition which may not be easy to distinguish from a
second order transition. We have checked that this is the case up
to the value $k=10.$


\section{Monte Carlo Results}

In this section we will study the gonihedric Potts model described
by Eq. (\ref{hamil1}) using Monte Carlo techniques. The phase
diagram of the model will obviously depend on $k$ and $q$. We
consider the model for the values $k=0, \ 0.5$ and $1.0$ and
$q>2$. In particular we will focus on $q=3,4,5.$ As we have
explained, the $q=2$ case is equivalent to the standard gonihedric
spin model, which has already been studied in the past. One of its
main results for $q=2$ has been that the phase transition is of
first order for small $k$ but it becomes of second order around
$k=0.5.$ The mean field approach suggests that for $q>2$ the phase
transition weakens, but it never becomes second order. We will
check this issue with the Monte Carlo method. The outcome is that
for $q>2$ the phase transition becomes of second order for quite
bigger values of $k.$ In addition, we find out that the value of
$k$ where the transition becomes second order depends on $q$ but
in any case it is bigger than $1.0$ for $q>2.$ This is the last
topic examined: we perform hysteresis loops for the values of $k:
1.0, 1.5, 2.0,$ for $q=3$ and consider the changing magnitude of
the loops.

We used a single flip Metropolis algorithm for the
updating and imposed periodic boundary conditions.

\subsection { The plaquette interaction case ($k=0$)}
\begin{figure}[!ht] \begin{center}
\includegraphics[scale=0.40,angle=-90]{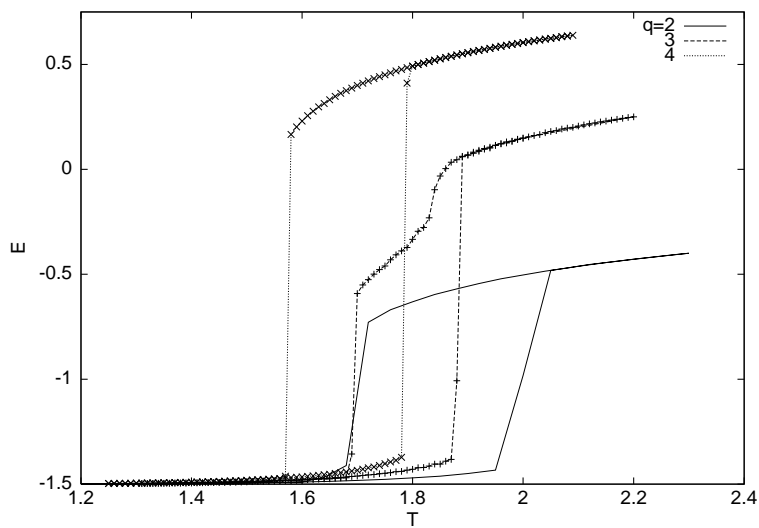}
\caption{Hysteresis loop results for $k=0$ and $q=2,3,4$ on a
$20^3$ lattice volume.} \label{fig3}
\end{center}
\end{figure}

For $k=0$ only the last term in equation (\ref{hamil1})
contributes, so the Potts model has just the plaquette
interaction. For the standard gonihedric model $(q=2)$ the system
has a strong first order phase transition
\cite{esbaijo},\cite{lip}. The mean field results suggest that the
order of the phase transition does not change for $q>2.$ To check
the validity of this prediction we have studied the system using
the  hysteresis loop method. In Fig. \ref{fig3} we present our
results for energy versus temperature on a $20^3$ lattice volume
for $q=3,4$.  For comparison reasons we also plot the well known
case for $q=2$. The unambiguous formation of large hysteresis
loops accompanied by large jumps in energy in the critical region
lead to a more or less safe conclusion of a first order phase
transition. The critical temperature $T_c$ takes smaller values as
$q$ increases and this feature is in  qualitative agreement  with
the mean field prediction, which may be seen in Fig. \ref{fig1}.

\subsection{The $k=0.5$ case}

For this value of $k$ all three terms in the Hamiltonian
contribute. We have chosen this value because, in the vicinity of
$k=0.5,$ the phase transition of the $q=2$ case  becomes second
order \cite{esbaijo}. In figure 4a we show the hysteresis loops
for the energy versus the temperature for $q=2,3,4,5$ and a $20^3$
lattice. The absence of any loop for $q=2$  agrees with previous
results indicating  a second order phase transition. However,
things change  for $q>2:$ it is evident that the loops for
$q=3,4,5$ grow bigger and the jump towards large values for the
energy is clear at least for the $q=4,5$ cases.  The conclusion suggested by the
results in Fig. 4a is that at $k=0.5$ the phase transition is
stronger for bigger $q.$ In Fig. 4b we show a two peak signal for
$q=3$ for  $10^3$ and $12^3$ lattice. The comparison for these two
lattice volumes shows that for the $12^3$ lattice the separation
between the two peaks is complete. This is a strong indication
that we have a first order transition.
\begin{figure}[!ht]
\subfigure[]{\includegraphics[scale=0.30,angle=-90]{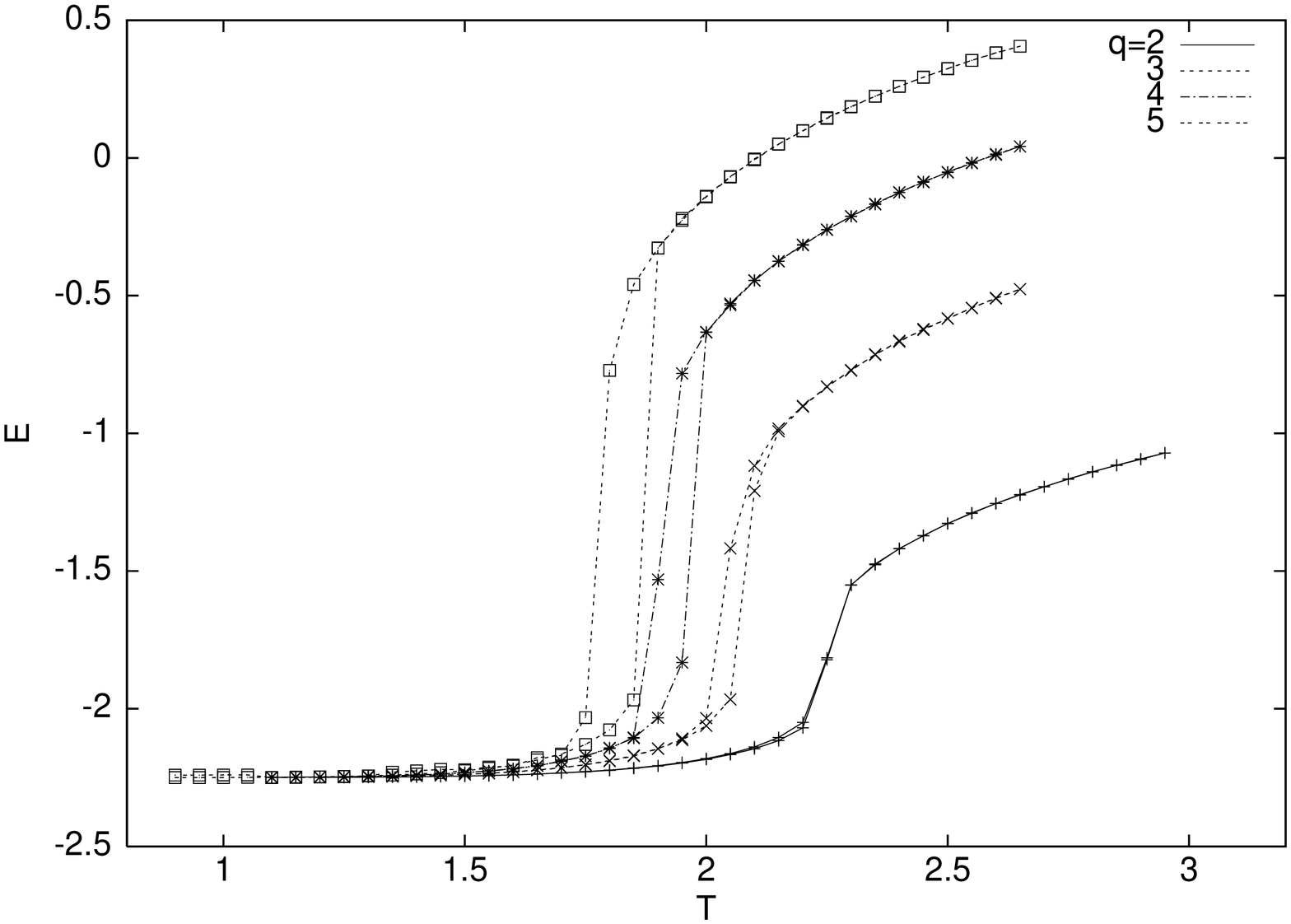}}
\subfigure[]{\includegraphics[scale=0.30,angle=-90]{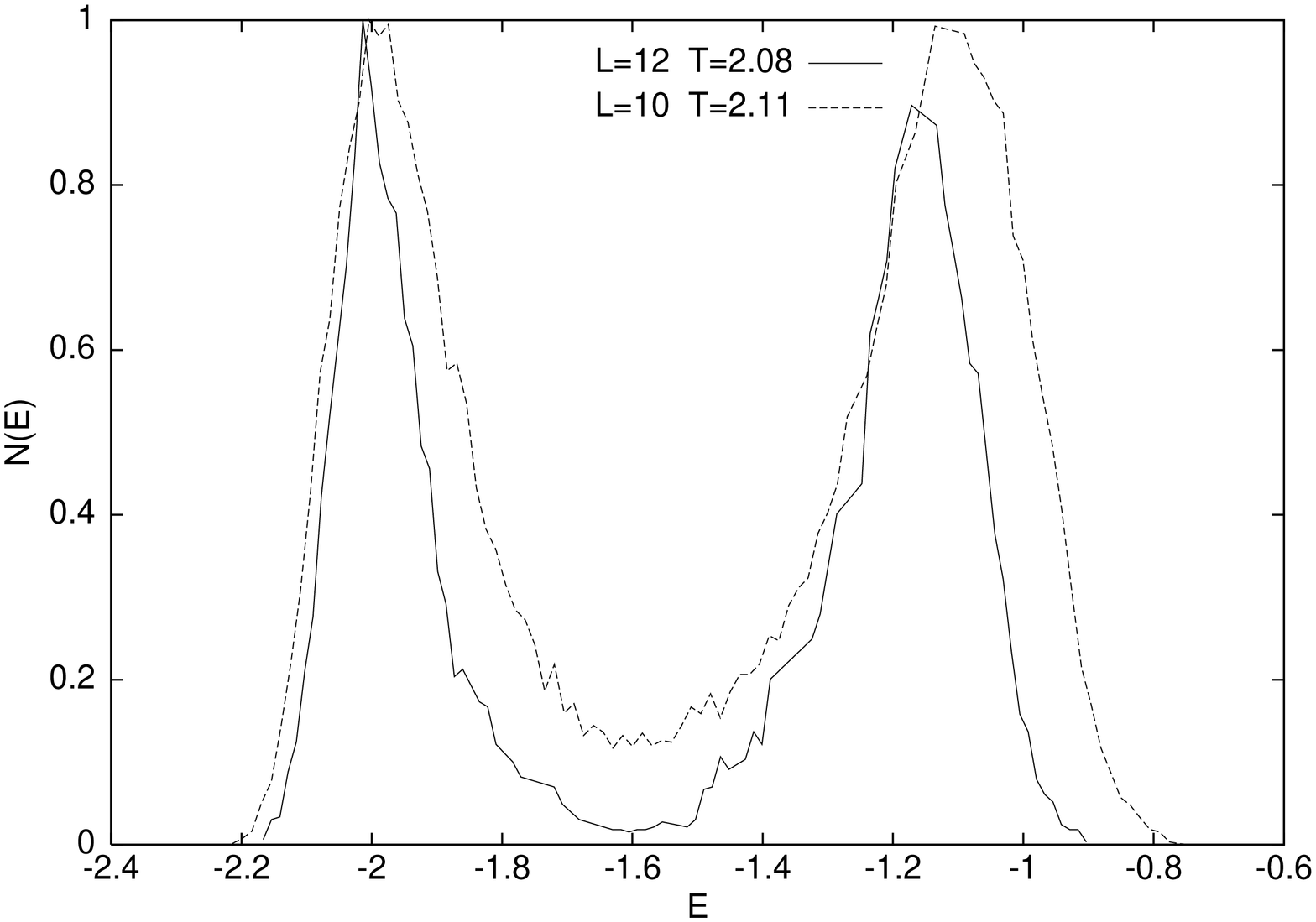}}
\caption{$k=0.5$: (a) Hysteresis loop results for $q=2,3,4,5$. (b)
A two peak signal for $q=3$  on $10^3$ and $12^3$ lattice
volumes.} \label{fig4ab} \end{figure}

\subsection{Nearest and next-to-nearest neighbour interaction ($k=1.0$)}

The plaquette interaction does not contribute in this case. A
similar model for $q=3$, has been studied in the past \cite{gavai,
billoire}, and found to have a first order phase transition. The
distinctive feature of the model we study here is that the ratio
of the ferromagnetic and the antiferromagnetic couplings is four;
this induces the huge degeneracy of the ground state that has been
described above and is not shared by the model in references
\cite{gavai,billoire}. It is of interest to detect if there are
any consequences of this peculiar feature of the model.

\begin{figure}[!ht]
\subfigure[]{\includegraphics[scale=0.30]{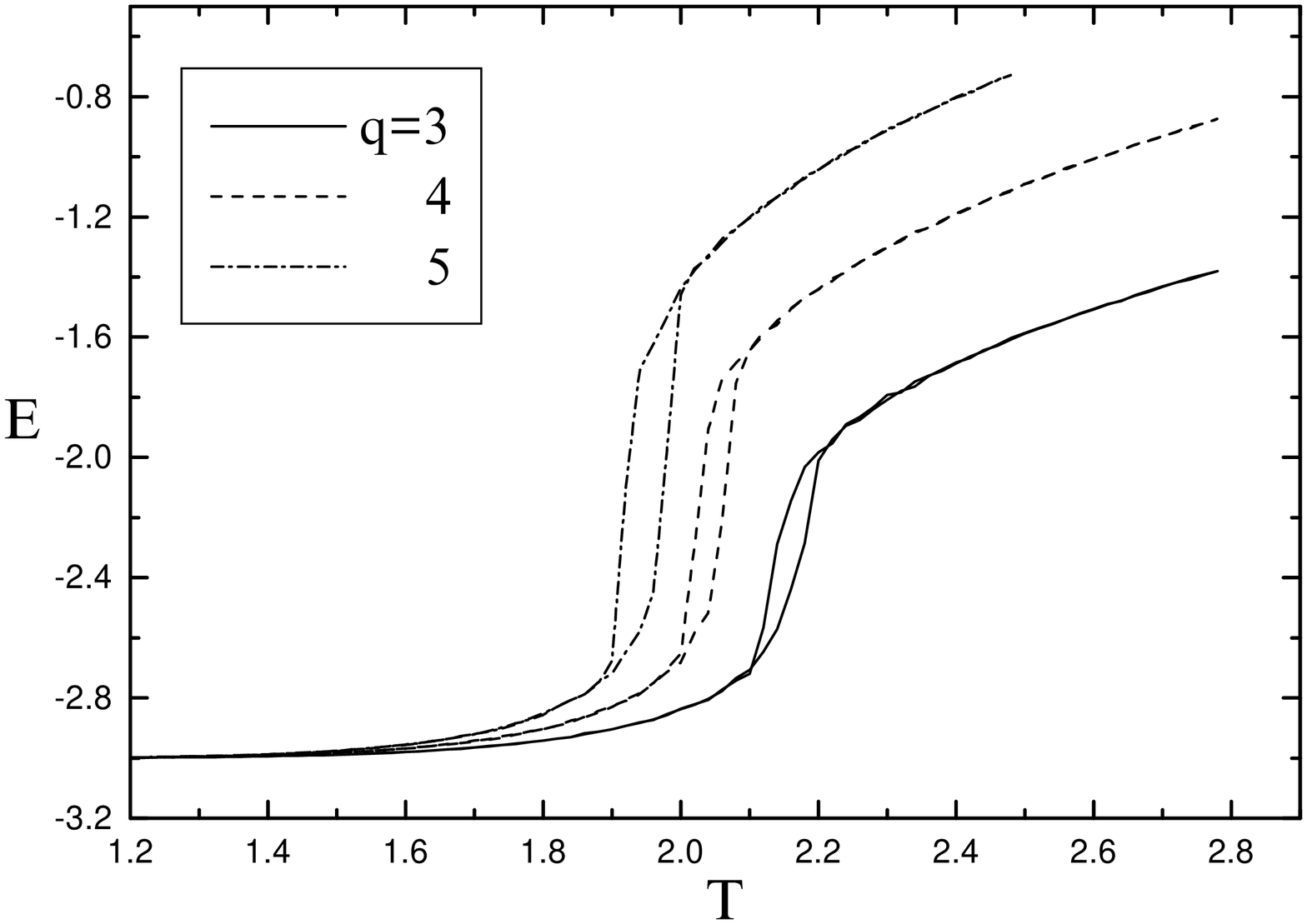}}
\subfigure[]{\includegraphics[scale=0.30]{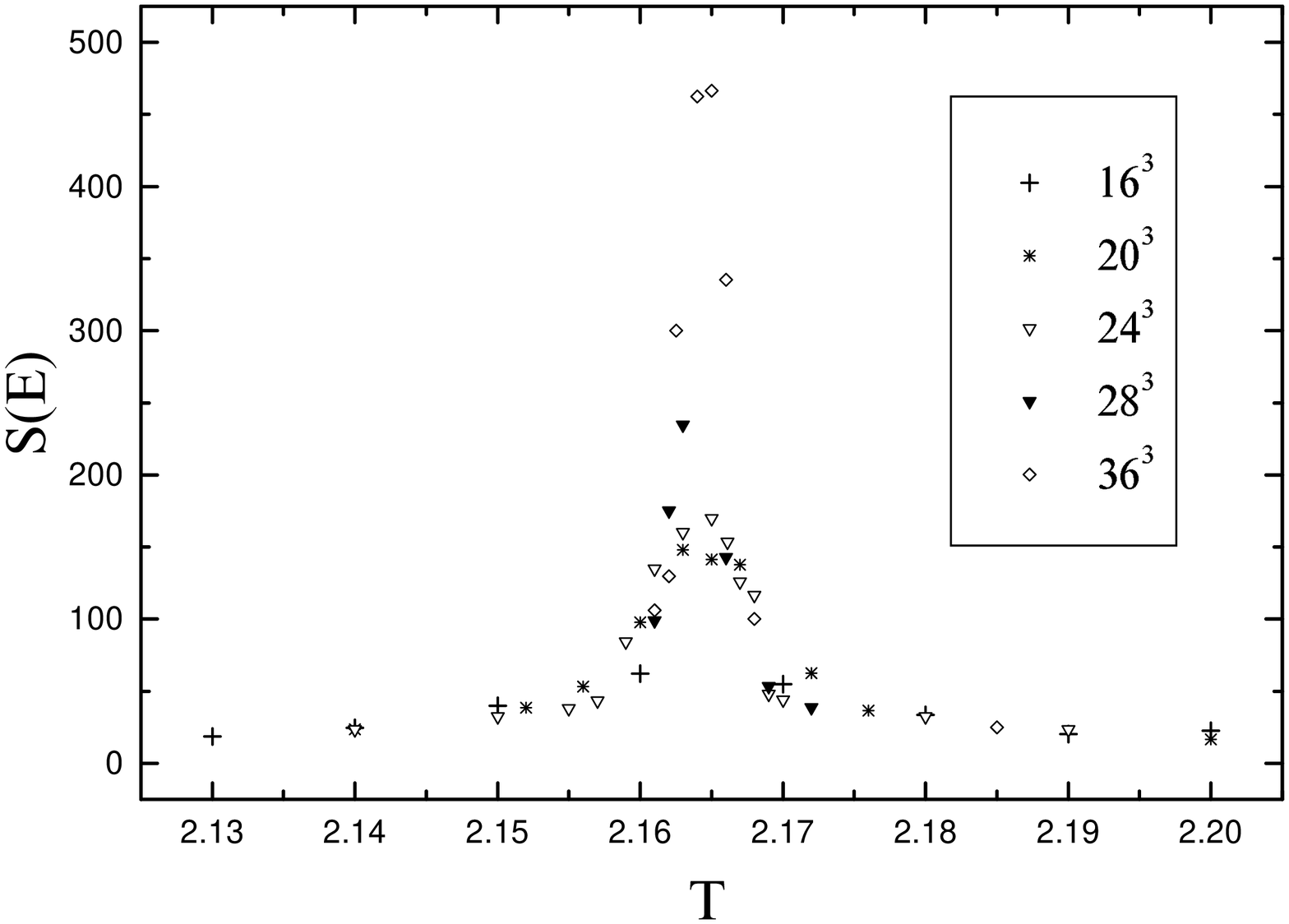}}
\caption{$k=1$: (a) Hysteresis loop results for $q=3,4,5$.
 (b) Susceptibility of the energy for five lattice volumes.(The statistical
errors are small enough and they are not included for the figure
not to be confused.)} \label{fig5ab}\end{figure}

The phase transition in the standard $q=2$ case is of second order
\cite{esbaijo}, \cite{dimop}. However, for bigger values of $q,$
the phase transition appears to be of first order. In Fig. 5a we
depict the hysteresis loops for $q=3,4,5$ in a $20^3$ lattice. The
clear loops for $q=4$ and $q=5$ give a strong indication for a
first order phase transition. The $q=3$ case is not so clear, so
in this case we also measure the susceptibility of the energy,
$S(E) \equiv V(<E^2>-<E>^2)$. The average energy has been measured
over  200000-800000 measurements, sampling every fourth iteration.
We have used five lattice volumes, namely $16^3, 20^3, 24^3, 28^3$
and $36^3$. The results depicted in Fig. 5b show that the peaks of
the susceptibility increase almost linearly with the volume which
is definitely the volume dependence characterizing a first order
phase transition.

\subsection{Larger values of $k$}

\begin{figure}[!h] \begin{center}
\includegraphics[scale=0.40,angle=-90]{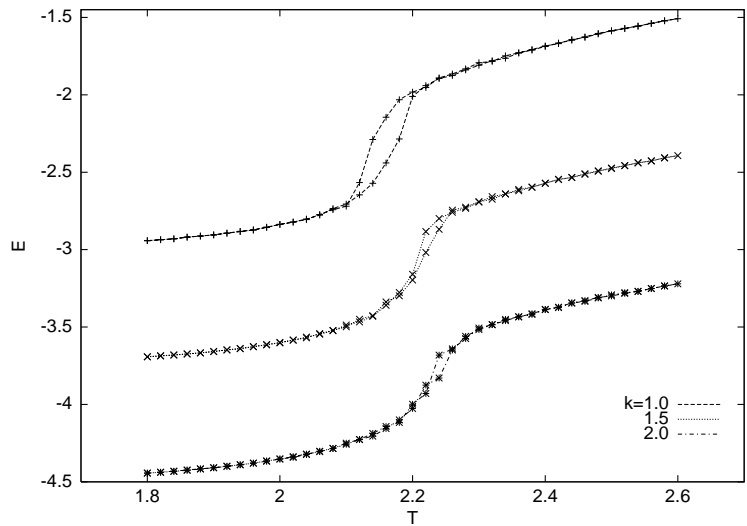}
\caption{Hysteresis loop results on $20^3$ lattice volume for
$q=3$ and $k=1.0, 1.5, 2.0.$} \label{fig6} \end{center}
\end{figure}

On the basis of the results of the previous paragraph one would
expect that the phase transition will possibly become of second
order at some  $k_c > 1.$ Since the phase transition grows
stronger with increasing $q,$ we expect that $k_c$ is an
increasing function of $q.$ The precise evaluation of $k_c(q)$
would require heavy numerical work. Nevertheless, we can achieve
qualitative understanding of the behaviour of the system. In Fig.
\ref{fig6} we depict the hysteresis loops for the $q=3$ Potts
model at $k=1.0, \ 1.5,$ and $2.0$ on a $20^3$ lattice volume. As
$k$ increases  the loop gradually disappears. Thus the transition
may become of second order at some value $k_c(3)$ greater than
$1.0.$ Recalling that $k_c(2) \simeq 0.5,$ it appears that the
hierarchy:
$$k_c(2) < k_c(3) < \dots.$$ might possibly be valid.

\vfill

\end{document}